\documentclass{aa}
\usepackage{color}
\usepackage{graphicx}
\usepackage{lscape}
\usepackage{natbib}
\usepackage[varg]{txfonts}
\usepackage{url}
\usepackage{xspace}
\bibpunct{(}{)}{;}{a}{}{,}
\newcounter{Rco}
\newcommand{\Ionst}[1]{\setcounter{Rco}{#1}\Roman{Rco}}

\newcommand{\Ionww}[3]{\mbox{#1\,{\scriptsize\Ionst{#2}}~$\lambda\lambda\,#3$\,\AA}\xspace}

\newcommand{\loggw}[1]{\mbox{$\log g\hspace{-0.5mm} =\hspace{-0.5mm}  #1$}}

\newcommand{\sga}{\raisebox{-0.10em}{$\stackrel{>}{{\mbox{\tiny $\sim$}}}$}}

\newcommand{\Teff}{\mbox{$T_\mathrm{eff}$}\xspace}
\newcommand{\Teffw}[1]{\mbox{$\Teff\hspace{-0.5mm} =\hspace{-0.5mm} #1 \,\mathrm{K}$}}

\newcommand{\gb}{\object{G191$-$B2B}\xspace}
\newcommand{\re}{\object{RE\,0503$-$289}\xspace}
\begin{document}
\title{Stellar laboratories}
\subtitle{II. New \ion{Zn}{iv} and \ion{Zn}{v} oscillator strengths \\
              and their validation in the hot white dwarfs \gb and \re
           \thanks
           {Based on observations with the NASA/ESA Hubble Space Telescope, obtained at the Space Telescope Science 
            Institute, which is operated by the Association of Universities for Research in Astronomy, Inc., under 
            NASA contract NAS5-26666.
           }
           \thanks
           {Based on observations made with the NASA-CNES-CSA Far Ultraviolet Spectroscopic Explorer.
           }
           \thanks
           {Tables 1 and 2 are only available at the CDS via anonymous ftp to
            cdsarc.u-strasbg.fr (130.79.128.5) or via
            http://cdsarc.u-strasbg.fr/viz-bin/qcat?J/A+A/.../...
           }
         }
\titlerunning{Stellar laboratories: new \ion{Zn}{iv} and \ion{Zn}{v} oscillator strengths}

\author{T\@. Rauch\inst{1}
        \and
        K\@. Werner\inst{1}
        \and
        P\@. Quinet\inst{2,3}
        \and
        J\@. W\@. Kruk\inst{4}
        }

\institute{Institute for Astronomy and Astrophysics,
           Kepler Center for Astro and Particle Physics,
           Eberhard Karls University,
           Sand 1,
           72076 T\"ubingen,
           Germany,
           \email{rauch@astro.uni-tuebingen.de}
           \and
           Astrophysique et Spectroscopie, Universit\'e de Mons -- UMONS, 7000 Mons, Belgium
           \and
           IPNAS, Universit\'e de Li\`ege, Sart Tilman, 4000 Li\`ege, Belgium
           \and
           NASA Goddard Space Flight Center, Greenbelt, MD\,20771, USA}

\date{Received 23 January 2014; accepted 3 March 2014}

\abstract {For the spectral analysis of high-resolution and high-signal-to-noise spectra of hot stars,
           state-of-the-art non-local thermodynamic equilibrium (NLTE) 
           model atmospheres are mandatory. These are strongly
           dependent on the reliability of the atomic data that is used for their calculation.
           In a recent analysis of the ultraviolet (UV) spectrum of the DA-type white dwarf \gb, 
           21 \ion{Zn}{iv} lines were newly identified. 
           Because of the lack of \ion{Zn}{iv} data, transition probabilities of the 
           isoelectronic \ion{Ge}{vi} were adapted for a first, coarse determination of the 
           photospheric Zn abundance.
          }
          {Reliable \ion{Zn}{iv} and \ion{Zn}{v} oscillator strengths 
           are used to improve the Zn abundance determination and to identify more
           Zn lines in the spectra of \gb and the DO-type white dwarf \re.
          }
          {We performed new calculations of \ion{Zn}{iv} and \ion{Zn}{v} oscillator strengths
           to consider their radiative and collisional bound-bound transitions
           in detail in our NLTE stellar-atmosphere models
           for the analysis of the \ion{Zn}{iv - v} spectrum exhibited in
           high-resolution and high-S/N UV observations of \gb  and \re.
          }
          {In the UV spectrum of \gb, we identify
              31 \ion{Zn}{iv} and 
              16 \ion{Zn}{v}  lines.
           Most of these are identified for the first time in any star.
           We can reproduce well almost all of them at 
           $\log \mathrm{Zn} = -5.52 \pm 0.2$ (mass fraction, about 1.7 times solar). 
           In particular, the \ion{Zn}{iv} / \ion{Zn}{v}
           ionization equilibrium, which is a very sensitive \Teff\ indicator, is well reproduced
           with the previously determined \Teffw{60\,000 \pm 2000} and \loggw{7.60 \pm 0.05}.
           In the spectrum of \re, we identified 128 \ion{Zn}{v} lines for the first time and determined
           $\log \mathrm{Zn} = -3.57 \pm 0.2$ (155 times solar).
          }
          {Reliable measurements and calculations of atomic data are a pre-requisite for
           stellar-atmosphere modeling. 
           Observed \ion{Zn}{iv} and \ion{Zn}{v} line profiles in two white dwarf (\gb and \re) 
           ultraviolet spectra were well reproduced with our newly calculated oscillator strengths. 
           This allowed us to determine the photospheric Zn abundance of these two stars precisely.
          }

\keywords{atomic data --
          line: identification --
          stars: abundances --
          stars: individual: \gb, \re\ --
          virtual observatory tools
         }

\maketitle

\section{Introduction}
\label{sect:intro}

In a recent spectral analysis of the hydrogen-rich DA-type white dwarf \gb, \citet{rauchetal2013} identified and reproduced
stellar lines of C, N, O, Al, Si, O, P, S, Fe, Ni, Ge, and Sn. In addition,
they identified 21 \ion{Zn}{iv} lines. The determined Zn abundance (logarithmic mass fraction of $-4.89$,
7.5 $\times$ solar) was uncertain because the unknown \ion{Zn}{iv} oscillator strengths were
approximated by values of the isoelectronic \ion{Ge}{vi} taken from \citet{rauchetal2012}.

In this paper, we introduce new oscillator strengths for \ion{Zn}{iv} and \ion{Zn}{v} (Sect.\,\ref{sect:zntrans}).
Then, we describe briefly our observations (Sect.\,\ref{sect:observation}),
our analysis strategy (Sect.\,\ref{sect:models}),
and revisit \gb to perform a precise determination of its Zn abundance (Sect.\,\ref{sect:model}).
The white dwarf \re is hotter than \gb and its trans-iron element abundances are strongly oversolar 
\citep{werneretal2012,
rauchetal2013} and, thus, it appears promising to identify Zn lines.
In Sect.\,\ref{sect:re}, we describe our search for these and the determination of its Zn abundance.
We summarize our results and conclude in Sect.\,\ref{sect:results}.

\section{Transition probabilities in \ion{Zn}{iv} and \ion{Zn}{v}}
\label{sect:zntrans}

\begin{figure}
   \resizebox{\hsize}{!}{\includegraphics{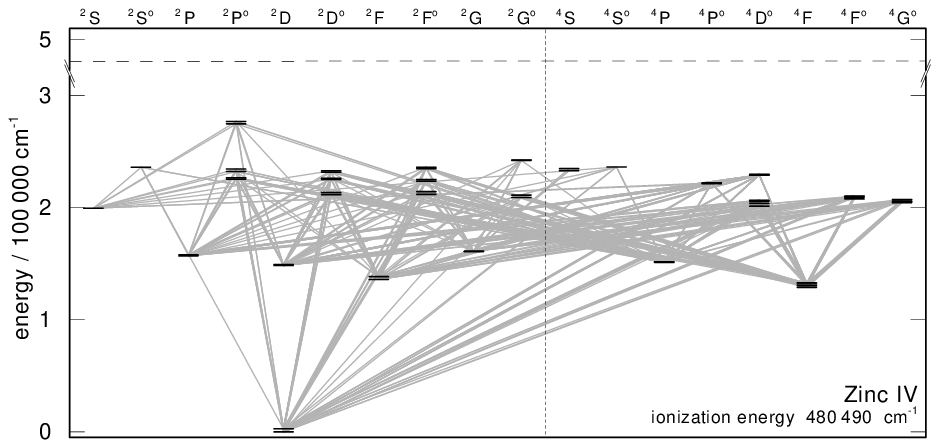}}
   \resizebox{\hsize}{!}{\includegraphics{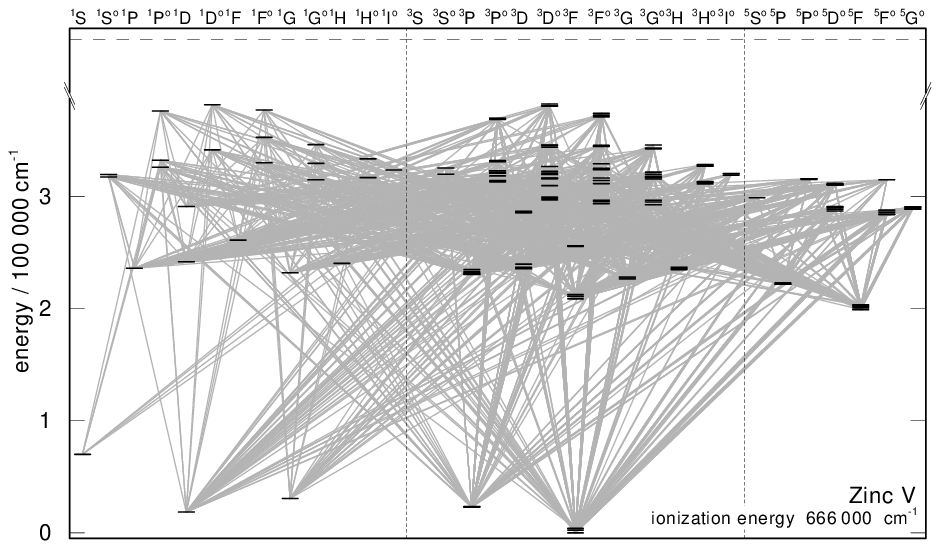}}
    \caption{Grotrian diagrams of our 
             \ion{Zn}{iv} (top) and
             \ion{Zn}{v} (bottom) model ions.
             Horizontal bars indicate levels,
             gray lines represent radiative
             transitions with known $f$ values, respectively.
             The dashed lines show the ionization energies.
            }
   \label{fig:grotrian}
\end{figure}

Radiative decay rates (oscillator strengths and transition probabilities) have been computed 
using the pseudo-relativistic Hartree-Fock (HFR) method as described by \citet{cowan1981}. 
For \ion{Zn}{iv}, configuration interaction has been considered among the configurations 
3d$^9$, 3d$^8$4s, 3d$^8$5s, 3d$^8$4d, 3d$^8$5d, 3d$^7$4s$^2$, 3d$^7$4p$^2$, 3d$^7$4d$^2$, 3d$^7$4f$^2$, 3d$^7$4s5s, 3d$^7$4s4d, and 3d$^7$4s5d 
for the even parity and 
3d$^8$4p, 3d$^8$5p, 3d$^8$4f, 3d$^8$5f, 3d$^7$4s4p, 3d$^7$4s5p, 3d$^7$4s4f, 3d$^7$4s5f, and 3d$^7$4p4d 
for the odd parity. Using experimental energy levels published by \citet{sugarmusgrove1995}, 
the average energies ($E_\mathrm{av}$), the Slater integrals ($F^\mathrm{k}$, $G^\mathrm{k}$), 
the spin-orbit parameters ($\zeta_\mathrm{nl}$), and the 
effective interaction parameters ($\alpha$, $\beta$) corresponding to 
3d$^9$, 3d$^8$4s, 3d$^8$4p
configurations were optimized 
using a well-established least-squares fitting process minimizing the differences between calculated 
and experimental energy levels within both configurations.  
     In the case of \ion{Zn}{v}, the configurations included in the HFR model were 
3d$^8$, 3d$^7$4s, 3d$^7$5s, 3d$^7$4d, 3d$^7$5d, 3d$^6$4s$^2$, 3d$^6$4p$^2$, 3d$^6$4d$^2$, 3d$^6$4s5s, 3d$^6$4s4d, 3d$^6$4s5d 
for the even parity and 
3d$^7$4p, 3d$^7$5p, 3d$^7$4f, 3d$^7$5f, 3d$^6$4s4p, 3d$^6$4s5p, 3d$^6$4s4f and 3d$^6$4p4d 
for the odd parity. For this ion, the semi-empirical fitting process was performed to optimize 
the radial integrals corresponding to 
3d$^8$, 3d$^7$4s, and 3d$^7$4p 
configurations using the experimental 
energy levels compiled by \citet{sugarmusgrove1995}. 
     The HFR oscillator strengths ($\log gf$) and transition probabilities ($gA$, in s$^{-1}$) for \ion{Zn}{iv} and \ion{Zn}{v}
spectral lines are reported in Tables \ref{tab:zniv:loggf} and \ref{tab:znv:loggf}, respectively, alongside with 
the numerical values (in cm$^{-1}$) 
of lower and upper energy levels and the corresponding wavelengths (in \AA). In the last column of each table, 
we also give the cancellation factor CF as defined by \citet{cowan1981}. We note that very small values of this 
factor (typically $< 0.05$) indicate strong cancellation effects in the calculation of line 
strengths. In these cases, the corresponding $gf$ and $gA$ values could be very inaccurate and so need 
to be considered with some care. However, very few transitions appearing in Tables \ref{tab:zniv:loggf} and 
\ref{tab:znv:loggf} are affected by these effects. 
Figure\,\ref{fig:grotrian} shows Grotrian diagrams of \ion{Zn}{iv} and \ion{Zn}{v} including
all levels and transitions from Tables \ref{tab:zniv:loggf} and \ref{tab:znv:loggf}.

\onltab{
\onecolumn
\begin{longtable}{rrccrrccrrr}
\caption{\label{tab:zniv:loggf}Calculated HFR oscillator strengths ($\log gf$) and transition probabilities (gA, in sec$^{-1}$) 
                               in \ion{Zn}{iv}. CF is the cancellation factor as defined by \citet{cowan1981}. 
                               In columns 3 and 6, e is written for even and o for odd.} \\
\hline\hline
\noalign{\smallskip}
                 & \multicolumn{3}{c}{Lower level} && \multicolumn{3}{c}{Upper level} &           &      &    \\
\cline{2-4}\cline{6-8}
\noalign{\smallskip}
Wavelength / \AA & Energy / cm$^{-1}$ & Parity & $j$  && Energy / cm$^{-1}$ & Parity & $j$  & $\log gf$ & $gA$ & CF \\
\noalign{\smallskip}
\hline
\endfirsthead
\caption{continued.}\\
\hline\hline
\noalign{\smallskip}
                 & \multicolumn{3}{c}{Lower level} && \multicolumn{3}{c}{Upper level} &           &      &    \\
\cline{2-4}\cline{6-8}
\noalign{\smallskip}
Wavelength / \AA & Energy / cm$^{-1}$ & Parity & $j$  && Energy / cm$^{-1}$ & Parity & $j$  & $\log gf$ & $gA$ & CF \\
\noalign{\smallskip}
\hline
\noalign{\smallskip}
\endhead
\hline
\noalign{\smallskip}
\endfoot
\noalign{\smallskip}
     361.161 &	       0 & e &	2.5 & &   276885 & o &  1.5 &$-$1.14 &  $3.71 \times 10^{09}$ & 0.417 \\
     364.796 &	    2759 & e &	1.5 & &   276885 & o &  1.5 &$-$1.98 &  $5.26 \times 10^{08}$ & 0.427 \\
     367.664 &	    2759 & e &	1.5 & &   274746 & o &  0.5 &$-$1.28 &  $2.59 \times 10^{09}$ & 0.471 \\
     412.677 &	       0 & e &	2.5 & &   242320 & o &  3.5 &$-$2.09 &  $3.18 \times 10^{08}$ & 0.399 \\
     423.415 &	       0 & e &	2.5 & &   236175 & o &  1.5 &$-$2.60 &  $9.35 \times 10^{07}$ & 0.446 \\
     423.533 &	       0 & e &	2.5 & &   236109 & o &  2.5 &$-$1.61 &  $9.07 \times 10^{08}$ & 0.159 \\
     425.890 &	       0 & e &	2.5 & &   234802 & o &  3.5 &$-$0.13 &  $2.71 \times 10^{10}$ & 0.638 \\
     428.420 &	    2759 & e &	1.5 & &   236175 & o &  1.5 &$-$3.53 &  $1.06 \times 10^{07}$ & 0.225 \\
\hline
\end{longtable}
\twocolumn
}

\onltab{
\onecolumn
\begin{longtable}{rrccrrccrrr}
\caption{\label{tab:znv:loggf}Like Table\,\ref{tab:zniv:loggf} for  \ion{Zn}{v}.} \\
\hline\hline
\noalign{\smallskip}
                 & \multicolumn{3}{c}{Lower level} && \multicolumn{3}{c}{Upper level} &           &      &    \\
\cline{2-4}\cline{6-8}
\noalign{\smallskip}
Wavelength / \AA & Energy / cm$^{-1}$ & Parity & $j$  && Energy / cm$^{-1}$ & Parity & $j$  & $\log gf$ & $gA$ & CF \\
\noalign{\smallskip}
\hline
\endfirsthead
\caption{continued.}\\
\hline\hline
\noalign{\smallskip}
                 & \multicolumn{3}{c}{Lower level} && \multicolumn{3}{c}{Upper level} &           &      &    \\
\cline{2-4}\cline{6-8}
\noalign{\smallskip}
Wavelength / \AA & Energy / cm$^{-1}$ & Parity & $j$  && Energy / cm$^{-1}$ & Parity & $j$  & $\log gf$ & $gA$ & CF \\
\noalign{\smallskip}
\hline
\noalign{\smallskip}
\endhead
\hline
\noalign{\smallskip}
\endfoot
\noalign{\smallskip}
\hline
     261.492 &	       0 & e &	4.0 & &	   382420 & o &	 3.0 &$-$1.35 &	 $4.37 \times 10^{09}$ & 0.084 \\
     263.190 &	    2466 & e &	3.0 & &	   382420 & o &	 3.0 &$-$2.37 &	 $4.12 \times 10^{08}$ & 0.074 \\
     263.710 &	    2466 & e &	3.0 & &	   381670 & o &	 2.0 &$-$2.40 &	 $3.81 \times 10^{08}$ & 0.062 \\
     264.246 &	    2466 & e &	3.0 & &	   380902 & o &	 2.0 &$-$1.48 &	 $3.17 \times 10^{09}$ & 0.103 \\
     264.282 &	    4036 & e &	2.0 & &	   382420 & o &	 3.0 &$-$3.98 &	 $1.01 \times 10^{07}$ & 0.024 \\
     264.807 &	    4036 & e &	2.0 & &	   381670 & o &	 2.0 &$-$2.68 &	 $2.01 \times 10^{08}$ & 0.034 \\
     265.151 &	       0 & e &	4.0 & &	   377144 & o &	 3.0 &$-$2.83 &	 $1.41 \times 10^{08}$ & 0.031 \\
     265.346 &	    4036 & e &	2.0 & &	   380902 & o &	 2.0 &$-$2.50 &	 $2.97 \times 10^{08}$ & 0.040 \\
\end{longtable}
\twocolumn}

\section{Observations}
\label{sect:observation}

In this analysis, we use the 
FUSE\footnote{Far Ultraviolet Spectroscopic Explorer} spectrum of \re and 
the FUSE and HST/STIS\footnote{Hubble Space Telescope / Space Telescope Imaging Spectrograph,
                  for our high-resolution spectrum of \gb, see \url{http://www.stsci.edu/hst/observatory/crds/calspec.html}} 
spectra \gb that are described in detail by \citet{werneretal2012} and \citet{rauchetal2013}, respectively.

Both FUSE spectra are co-added from all available observations of \re and \gb. They cover the wavelength range 
$910\,\mathrm{\AA} < \lambda <  1188\,\mathrm{\AA}$. Their resolving power is $R = \lambda/\Delta\lambda \approx 20\,000$.
The HST/STIS spectrum of \gb is co-added from 105 observation with the highest resolution (grating E140H,
$R \approx 118\,000$, $1145\,\mathrm{\AA} < \lambda <  1700\,\mathrm{\AA}$) available via MAST.

\section{Model atmospheres and atomic data}
\label{sect:models}

To determine the Zn abundance of \gb, it would be straightforward to use the
final model of \citet{rauchetal2013} as well as their model atoms with the only exception
that the \ion{Zn}{iv} and \ion{Zn}{v} model ions were replaced by the extended versions 
that consider the newly calculated transition probabilities (Sect.\,\ref{sect:zntrans}).
Unfortunately, the employed T\"ubingen NLTE model-atmosphere package 
\citep[][TMAP\footnote{\url{http://astro.uni-tuebingen.de/~TMAP}}]{werneretal2003,rauchdeetjen2003},
which is used to calculate plane-parallel, chemically homogeneous, metal-line blanketed NLTE model 
atmospheres, overcharged our FORTRAN compilers. The program would not compile if the array
sizes were increased further according to the much higher number of atomic levels 
treated in NLTE and the respective higher number of radiative and collisional transitions. 

Thus, we decided to reduce the number of N and O levels treated in NLTE 
(Table\,\ref{tab:statistics}) to create a TMAP executable. Test calculations have shown that 
the deviations in temperature and density structure between the final model of \citet{rauchetal2013} 
and a model with reduced N and O model atoms are negligible. 
Then, the Zn occupation numbers are determined in a line-formation calculation, i.e., at fixed 
temperature and density structure. Since Zn opacities were already considered in our start model,
the atmospheric structure and the background opacities are well modeled.

\begin{table}\centering
\caption{Statistics of our N, O, and Zn model atoms for \gb.}
\label{tab:statistics}
\begin{tabular}{r@{~}lrrr}
\hline\hline
\noalign{\smallskip}
\multicolumn{2}{c}{ion~~~~}  & NLTE levels &   LTE levels &  lines \\
\noalign{\smallskip}                                                        
\hline                                                                      
N  & {\sc ii}   &      1\hspace{5mm}\hbox{} & 246\hspace{4mm}\hbox{} &    0\hspace{2mm}\hbox{} \\
   & {\sc iii}  &      9\hspace{5mm}\hbox{} &  57\hspace{4mm}\hbox{} &   10\hspace{2mm}\hbox{} \\
   & {\sc iv}   &      9\hspace{5mm}\hbox{} &  85\hspace{4mm}\hbox{} &   10\hspace{2mm}\hbox{} \\
   & {\sc v}    &     10\hspace{5mm}\hbox{} &  52\hspace{4mm}\hbox{} &   20\hspace{2mm}\hbox{} \\
   & {\sc vi}   &      1\hspace{5mm}\hbox{} &   0\hspace{4mm}\hbox{} &    0\hspace{2mm}\hbox{} \\
\noalign{\smallskip}                                                 
O  & {\sc ii}   &      1\hspace{5mm}\hbox{} &  46\hspace{4mm}\hbox{} &    0\hspace{2mm}\hbox{} \\
   & {\sc iii}  &      9\hspace{5mm}\hbox{} &  63\hspace{4mm}\hbox{} &    6\hspace{2mm}\hbox{} \\
   & {\sc iv}   &      9\hspace{5mm}\hbox{} &  85\hspace{4mm}\hbox{} &   11\hspace{2mm}\hbox{} \\
   & {\sc v}    &      9\hspace{5mm}\hbox{} & 117\hspace{4mm}\hbox{} &   12\hspace{2mm}\hbox{} \\
   & {\sc vi}   &     10\hspace{5mm}\hbox{} &  75\hspace{4mm}\hbox{} &   19\hspace{2mm}\hbox{} \\
   & {\sc vii}  &      1\hspace{5mm}\hbox{} &   0\hspace{4mm}\hbox{} &    0\hspace{2mm}\hbox{} \\
\noalign{\smallskip}                                                  
Zn & {\sc ii}   &      6\hspace{5mm}\hbox{} &   0\hspace{4mm}\hbox{} &    8\hspace{2mm}\hbox{} \\
   & {\sc iii}  &     13\hspace{5mm}\hbox{} &   0\hspace{4mm}\hbox{} &   17\hspace{2mm}\hbox{} \\
   & {\sc iv}   &     63\hspace{5mm}\hbox{} &  13\hspace{4mm}\hbox{} &  399\hspace{2mm}\hbox{} \\
   & {\sc v}    &    157\hspace{5mm}\hbox{} &   0\hspace{4mm}\hbox{} & 1878\hspace{2mm}\hbox{} \\
   & {\sc vi}   &      1\hspace{5mm}\hbox{} &   0\hspace{4mm}\hbox{} &    0\hspace{2mm}\hbox{} \\
\hline
\end{tabular}
\end{table}

All model atoms (including Zn) are provided via the T\"ubingen Model-Atom Database
\citep[TMAD\footnote{\url{http://astro.uni-tuebingen.de/~TMAD}},][]{rauchdeetjen2003},
that has been set up within a project of the German Astrophysical Virtual Observatory 
(GAVO\footnote{\url{http://www.g-vo.org}}).
All SEDs that were calculated for this analysis are available via
the registered Theoretical Stellar Spectra Access 
(TheoSSA\footnote{\url{http://dc.g-vo.org/theossa}}) VO service.

\section{The photospheric Zn abundance in \gb}
\label{sect:model}

\begin{figure}
   \resizebox{\hsize}{!}{\includegraphics{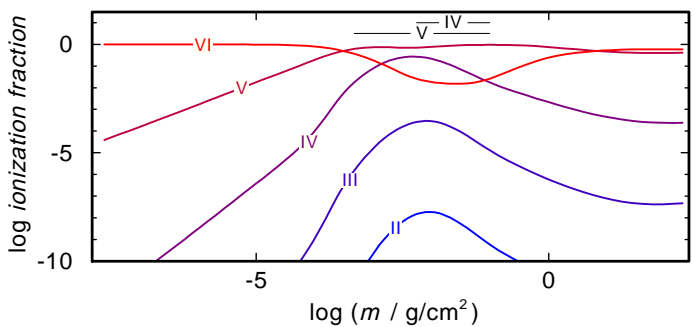}}
    \caption{Ionization fractions of Zn\,{\sc ii} - {\sc vi} in our \gb model atmosphere.
             $m$ is the column mass, measured from the outer boundary of our model atmosphere.
             The formation depths 
             (i.e\@., $\tau = 1$) 
             of the \ion{Zn}{iv} - {\sc v}
             line cores are marked.
            }
   \label{fig:ionization}
\end{figure}

\ion{Zn}{iv} and \ion{Zn}{v} are the dominant ionization stages of Zn in the atmosphere
of \gb (Fig.\,\ref{fig:ionization}). 
Therefore, we closely inspected the available spectra for lines of these ions.

\begin{figure*}
   \resizebox{\hsize}{!}{\includegraphics{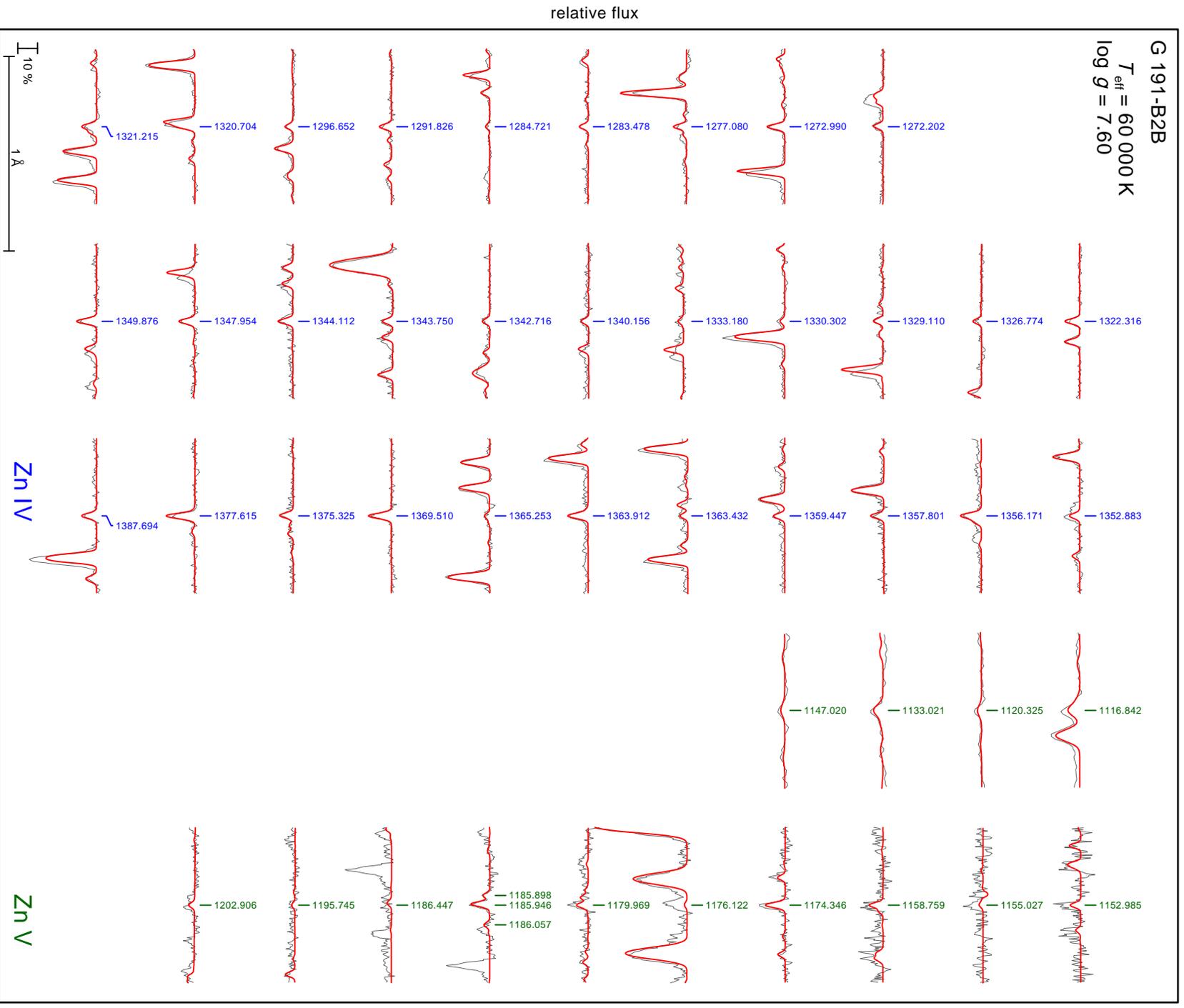}}
    \caption{\ion{Zn}{iv} lines (left  panel, marked with their wavelengths in \AA, blue in the online version) and 
             \ion{Zn}{v}  lines (right panel, marked in green)
             in the 
             FUSE (for lines at $\lambda < \mathrm{1150\,\AA}$) and 
             HST/STIS ($\lambda > \mathrm{1150\,\AA}$) observations of \gb
             compared with our theoretical line profiles. 
             For the identification of other lines, see \citet{rauchetal2013}.
             The vertical bar shows 10\,\% of the continuum flux.
            }
   \label{fig:znlines}
\end{figure*}

In the FUSE and HST/STIS 
observations of \gb \citep[cf\@.][]{rauchetal2013}
we identified
31 \ion{Zn}{iv} (10 new identifications) and
16 \ion{Zn}{v}  (all new) lines. 
The observed wavelength positions 
\citep[a radial velocity of $v_\mathrm{rad} = 22.1\,\mathrm{km/s}$ was applied according to][]{holbergetal1994, rauchetal2013}
deviate partly from those given in Tables \ref{tab:zniv:loggf}
and \ref{tab:znv:loggf} by some hundredths of an \AA. The good agreement of the strongest, unshifted
lines in our model\footnote{All synthetic spectra shown in this paper are convolved with Gaussians 
to match the spectral resolution (FUSE: FWHM = 0.06\,\AA, STIS: FWHM = 0.01\,\AA).}
(Fig.\,\ref{fig:znlines}) with the observations permits to shift the lines
to observed absorption features in their closest vicinity. The reason for this uncertainty is
most likely the limited accuracy of the \ion{Zn}{iv} and \ion{Zn}{v} energy levels from which
the wavelengths of the line transitions were calculated.
The identified lines are summarized in Table\,\ref{tab:lineids}.

\begin{longtab}
\begin{longtable}{rlr@{.}lr@{.}lp{34mm}}
\caption{\label{tab:lineids}Identified Zn lines in the UV spectrum of \gb. Observed wavelengths are given only 
                            in case that they deviate from the theoretical wavelengths (cf\@. Tables \ref{tab:zniv:loggf}
                            and \ref{tab:znv:loggf}). 
                            ``prev. unid.'' denotes lines that were previously listed as unidentified by \citet{rauchetal2013}.} \\
\hline\hline
\multicolumn{2}{c}{}         & \multicolumn{4}{c}{wavelength / \AA}                           &                          \\
\cline{3-6}
\multicolumn{7}{c}{}                                                                                     \vspace{-5.5mm} \\
\multicolumn{2}{c}{ion~~~~}  & \multicolumn{4}{c}{}                                           & comments \vspace{-5.0mm} \\
\multicolumn{2}{c}{}         & \multicolumn{2}{c}{theoretical} & \multicolumn{2}{c}{observed} &                          \\
\hline
\endfirsthead
\caption{continued.}\\
\hline\hline
\multicolumn{2}{c}{}         & \multicolumn{4}{c}{wavelength / \AA}                           &                          \\
\cline{3-6}
\multicolumn{7}{c}{}                                                                                     \vspace{-5.5mm} \\
\multicolumn{2}{c}{ion~~~~}  & \multicolumn{4}{c}{}                                           & comments \vspace{-5.0mm} \\
\multicolumn{2}{c}{}         & \multicolumn{2}{c}{theoretical} & \multicolumn{2}{c}{observed} &                          \\
\hline
\endhead
\hline
\noalign{\smallskip}
\endfoot
Zn & {\sc iv} & 1272&202 & \multicolumn{2}{c}{} &                                      \\
   &          & 1272&990 & 1272&975             & prev\@. unid\@.                      \\
   &          & 1277&080 & 1277&130             & $\Delta\lambda > 0.02\,\mathrm{\AA}$ \\
   &          & 1283&478 & 1283&525             & $\Delta\lambda > 0.02\,\mathrm{\AA}$, prev\@. unid\@. \\
   &          & 1284&721 & 1284&740             &                                      \\                         
   &          & 1291&826 & 1291&810             &                                      \\                         
   &          & 1296&652 & 1296&620             &                                      \\                         
   &          & 1320&704 & 1320&725             &                                      \\                         
   &          & 1321&215 & \multicolumn{2}{c}{} &                                      \\                         
   &          & 1322&316 & 1322&320             &                                      \\                         
   &          & 1326&774 & 1326&735             & $\Delta\lambda > 0.02\,\mathrm{\AA}$ \\
   &          & 1329&110 & \multicolumn{2}{c}{} &                                      \\                         
   &          & 1330&302 & 1330&325             &                                      \\                         
   &          & 1333&180 & \multicolumn{2}{c}{} &                                      \\                         
   &          & 1340&156 & 1340&190             & $\Delta\lambda > 0.02\,\mathrm{\AA}$ \\
   &          & 1342&716 & 1342&755             & $\Delta\lambda > 0.02\,\mathrm{\AA}$ \\
   &          & 1343&750 & 1343&815             & $\Delta\lambda > 0.02\,\mathrm{\AA}$ \\
   &          & 1344&122 & 1344&090             & $\Delta\lambda > 0.02\,\mathrm{\AA}$ \\
   &          & 1347&954 & 1347&970             &                                      \\                         
   &          & 1349&876 & 1349&895             &                                      \\                         
   &          & 1352&883 & 1352&905             & $\Delta\lambda > 0.02\,\mathrm{\AA}$ \\
   &          & 1356&171 & 1356&090             & $\Delta\lambda > 0.02\,\mathrm{\AA}$,
                                                  blend with \ion{Zn}{iv} $\lambda$ 1356.195\,\AA \\
   &          & 1357&801 & 1357&810             &                                      \\                         
   &          & 1359&477 & 1359&490             &                                      \\                         
   &          & 1363&432 & 1363&420             &                                      \\                         
   &          & 1363&912 & 1363&940             & $\Delta\lambda > 0.02\,\mathrm{\AA}$ \\
   &          & 1365&253 & 1365&260             &                                      \\                         
   &          & 1369&510 & 1369&515             &                                      \\                         
   &          & 1375&325 & \multicolumn{2}{c}{} &                                      \\                         
   &          & 1377&615 & 1377&635             &                                      \\                         
   &          & 1387&694 & 1387&720             &                                      \\                         
Zn & {\sc v}  & 1116&842 & 1116&860             &                                      \\                         
   &          & 1120&325 & 1120&330             &                                      \\                         
   &          & 1133&031 & 1133&060             & $\Delta\lambda > 0.02\,\mathrm{\AA}$, prev\@. unid\@. \\ 
   &          & 1147&020 & 1147&040             &                                      \\                         
   &          & 1152&985 & 1152&980             &                                      \\                         
   &          & 1155&027 & 1155&045             &                                      \\                         
   &          & 1158&759 & 1158&750             & prev\@. unid\@.                      \\                         
   &          & 1174&346 & 1174&325             & prev\@. unid\@.                      \\                         
   &          & 1176&122 & \multicolumn{2}{c}{} & weak, prev\@. unid\@.                \\                         
   &          & 1179&969 & 1180&005             & $\Delta\lambda > 0.02\,\mathrm{\AA}$ \\ 
   &          & 1185&898 & 1185&905             &                                      \\ 
   &          & 1185&948 & 1185&955             &                                      \\ 
   &          & 1186&057 & \multicolumn{2}{c}{} &                                      \\ 
   &          & 1186&447 & \multicolumn{2}{c}{} & weak, prev\@. unid\@.                \\ 
   &          & 1195&745 & \multicolumn{2}{c}{} & weak                                 \\ 
   &          & 1202&906 & \multicolumn{2}{c}{} &                                      \\ 
\hline
\end{longtable}
\end{longtab}

\begin{figure}
  \resizebox{\hsize}{!}{\includegraphics{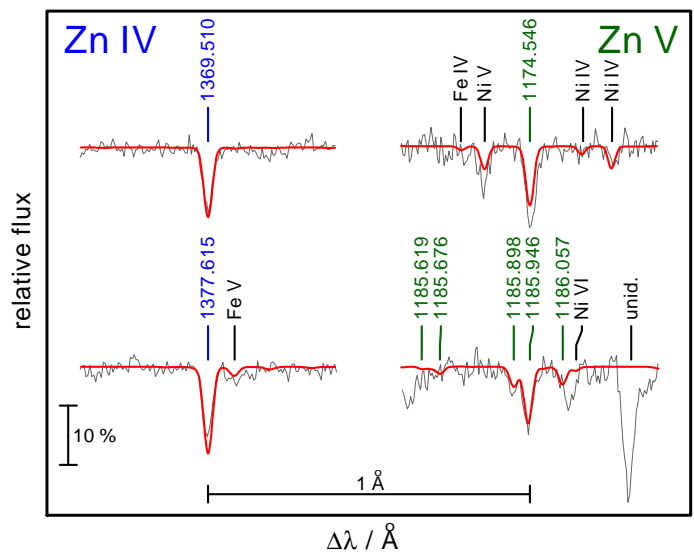}}
  \caption{Theoretical line profiles 
           of the strongest \ion{Zn}{iv} lines (left) and \ion{Zn}{v} lines (right)
           (marked with their wavelengths from Tables \ref{tab:zniv:loggf} and \ref{tab:znv:loggf})
           calculated from our model of \gb
           with a Zn abundance of $3.0 \times 10^{-6}$ (mass fraction)
           located in the STIS wavelength range compared with the observation.
           The lines are shifted to the observation, see Table \ref{tab:lineids}.
          } 
  \label{fig:znquinet}
\end{figure}

Our calculations have shown that the \ion{Zn}{iv}\,/\,\ion{Zn}{v} ionization equilibrium
at \Teffw{60\,000} and 
\loggw{7.6}
\citep[cf\@.][]{rauchetal2013} is well reproduced (Figs.\,\ref{fig:znlines}, \ref{fig:znquinet}).
On the other hand, the Zn abundance given by \citet[$1.3 \times 10^{-5} \pm 0.5\,\mathrm{dex}$ by mass]{rauchetal2013} 
is too high. We reduced it to $3.0 \times 10^{-6} \pm 0.2\,\mathrm{dex}$ 
\citep[about 1.7 times solar, following][]{asplundetal2009}
to reproduce the observed Zn lines best. This agrees with the previous value within the error limits.
Even a solar Zn abundance, however, is possible within the error limits.

\section{The photospheric Zn abundance in \re}
\label{sect:re}

\begin{figure}
  \resizebox{\hsize}{!}{\includegraphics{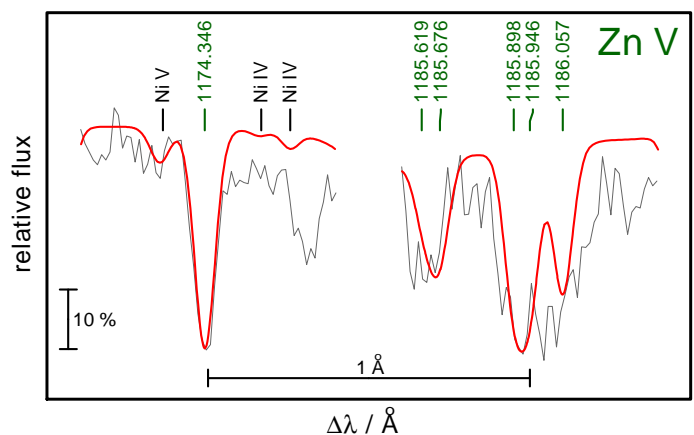}}
  \caption{Theoretical line profiles 
           of the strongest \ion{Zn}{v} lines 
           calculated from our model of \re
           with a Zn abundance of $2.7 \times 10^{-4}$ (mass fraction)
           located in the FUSE wavelength range compared with the observation.
           The lines are shifted to the observation, see Table \ref{tab:lineidsre}.
          } 
  \label{fig:znre0503}
\end{figure}

\begin{figure*}
   \resizebox{\hsize}{!}{\includegraphics{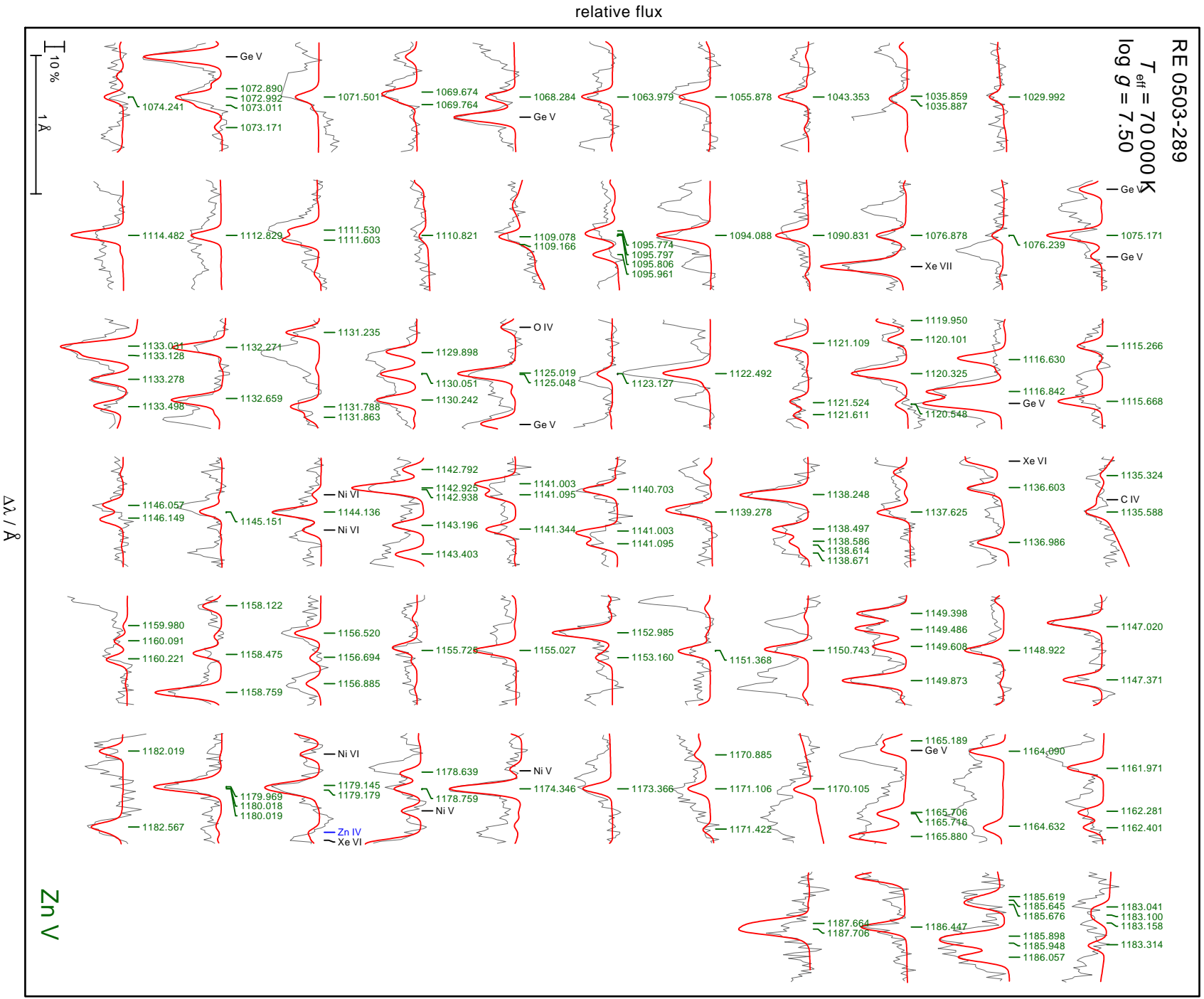}}
    \caption{\ion{Zn}{v} lines 
             in the 
             FUSE observation of \re
             compared with our theoretical line profiles. 
            }
   \label{fig:znlinesre}
\end{figure*}

Our inspection of all UV spectra that were already used by \citet{werneretal2012} has shown that 
only \re exhibits prominent Zn lines 
(Fig.\,\ref{fig:znre0503}). 
We find that in its FUSE spectrum ($v_\mathrm{rad} = 26.3\,\mathrm{km/s}$),
a rich \ion{Zn}{v} spectrum of 128 lines (Fig.\,\ref{fig:znlinesre}, Table\,\ref{tab:lineidsre}) is present. 
The synthetic spectrum of our final model shows more, weak lines that do not 
have an unambiguous line identification due to the S/N of the observation.
The model's prediction of their relative line strengths facilitates to distinguish between
noise and ``real'' lines in the observation and, hence, to identify even such weak lines.
In general, all lines with oscillator strengths $gf\,\sga\,0.01$ can be detected.

\begin{longtab}
\begin{longtable}{rlr@{.}lr@{.}lp{34mm}}
\caption{\label{tab:lineidsre}Like Table\,\ref{tab:lineids}, for \re.} \\
\hline\hline
\multicolumn{2}{c}{}         & \multicolumn{4}{c}{wavelength / \AA}                           &                         \\
\cline{3-6}
\multicolumn{7}{c}{}                                                                                     \vspace{-5.5mm} \\
\multicolumn{2}{c}{ion~~~~}  & \multicolumn{4}{c}{}                                           & comments \vspace{-5.0mm} \\
\multicolumn{2}{c}{}         & \multicolumn{2}{c}{theoretical} & \multicolumn{2}{c}{observed} &                          \\
\hline
\endfirsthead
\caption{continued.}\\
\hline\hline
\multicolumn{2}{c}{}         & \multicolumn{4}{c}{wavelength / \AA}                           &                         \\
\cline{3-6}
\multicolumn{7}{c}{}                                                                                     \vspace{-5.5mm} \\
\multicolumn{2}{c}{ion~~~~}  & \multicolumn{4}{c}{}                                           & comments \vspace{-5.0mm} \\
\multicolumn{2}{c}{}         & \multicolumn{2}{c}{theoretical} & \multicolumn{2}{c}{observed} &                          \\
\hline
\endhead
\hline
\noalign{\smallskip}
\endfoot
Zn & {\sc v}  &	1029&992 & \multicolumn{2}{c}{} &			               \\
   &          &	1035&859 & \multicolumn{2}{c}{} &		    		       \\
   &          &	1035&887 & \multicolumn{2}{c}{}	&		    		       \\
   &          &	1043&353 & \multicolumn{2}{c}{}	&		    		       \\
   &          &	1055&878 & \multicolumn{2}{c}{}	&		    		       \\
   &          &	1063&979 & \multicolumn{2}{c}{}	&		    		       \\
   &          &	1068&284 & \multicolumn{2}{c}{}	&		    		       \\
   &          &	1069&674 & \multicolumn{2}{c}{}	& blend with \ion{O}{v}, \ion{Ga}{v}, \ion{Ge}{v} \\
   &          &	1069&764 & \multicolumn{2}{c}{} &		    		       \\
   &          &	1071&501 & \multicolumn{2}{c}{}	&		    		       \\
   &          &	1072&992 & 1072&950	        & $\Delta\lambda > 0.02\,\mathrm{\AA}$ \\
   &          &	1074&241 & 1074&265	        & $\Delta\lambda > 0.02\,\mathrm{\AA}$ \\
   &          &	1075&171 & 1075&050	        & $\Delta\lambda > 0.02\,\mathrm{\AA}$ \\
   &          &	1076&239 & \multicolumn{2}{c}{}	&		    		       \\
   &          &	1076&878 & 1076&895	        &		    		       \\
   &          &	1090&831 & 1090&800	        & $\Delta\lambda > 0.02\,\mathrm{\AA}$ \\
   &          &	1094&088 & 1094&110	        & blend with \ion{N}{iv}               \\
   &          &	1095&774 & \multicolumn{2}{c}{}	& blend with \ion{Ge}{v}               \\
   &          &	1095&797 & \multicolumn{2}{c}{}	&		    		       \\
   &          &	1095&961 & 1095&945	        &		    		       \\
   &          &	1109&078 & 1109&110             & $\Delta\lambda > 0.02\,\mathrm{\AA}$ \newline
                                                  blend with \ion{C}{iv}               \\
   &          &	1109&166 & \multicolumn{2}{c}{}	& blend with \ion{C}{iv}               \\
   &          &	1110&821 & 1110&810	        & weak 	    		               \\
   &          &	1111&530 & \multicolumn{2}{c}{}	& blend with \ion{C}{iii}, \ion{O}{iv} \\
   &          &	1111&603 & \multicolumn{2}{c}{}	&		    		       \\
   &          &	1112&829 & \multicolumn{2}{c}{}	&		    		       \\
   &          &	1114&482 & \multicolumn{2}{c}{}	&		    		       \\
   &          &	1115&266 & 1115&295             & $\Delta\lambda > 0.02\,\mathrm{\AA}$ \\
   &          &	1115&668 & 1115&695             & $\Delta\lambda > 0.02\,\mathrm{\AA}$ \\
   &          &	1116&630 & \multicolumn{2}{c}{}	& too strong in model                  \\
   &          &	1116&842 & 1116&860             &  				       \\
   &          &	1119&950 & 1119&940             &  				       \\
   &          &	1120&101 & 1120&080             & $\Delta\lambda > 0.02\,\mathrm{\AA}$ \\
   &          &	1120&325 & \multicolumn{2}{c}{} &  				       \\
   &          &	1121&109 & 1121&095	        &		    		       \\
   &          &	1121&524 & \multicolumn{2}{c}{}	& weak	    		               \\
   &          &	1122&502 & \multicolumn{2}{c}{}	& blend with \ion{Si}{iv}              \\
   &          &	1123&127 & \multicolumn{2}{c}{}	& blend with \ion{Ga}{v}               \\
   &          &	1125&019 & 1125&050	        & $\Delta\lambda > 0.02\,\mathrm{\AA}$ \\
   &          &	1125&048 & 1125&060             &                                      \\
   &          &	1129&898 & \multicolumn{2}{c}{}	& blend with \ion{Ga}{v}	       \\
   &          &	1130&051 & \multicolumn{2}{c}{} &  				       \\
   &          &	1130&242 & \multicolumn{2}{c}{}	&		    		       \\
   &          &	1131&242 & 1131&250	        & prev. unid.                          \\
   &          &	1131&788 & \multicolumn{2}{c}{}	&		    		       \\
   &          &	1131&863 & \multicolumn{2}{c}{}	&		    		       \\
   &          &	1132&271 & 1132&290	        &		    		       \\
   &          &	1132&659 & \multicolumn{2}{c}{}	& blend with \ion{N}{iv}	       \\
   &          &	1133&031 & 1133&060             & $\Delta\lambda > 0.02\,\mathrm{\AA}$ \\
   &          &	1133&128 & \multicolumn{2}{c}{}	&		    		       \\
   &          &	1133&278 & 1133&300  	        & $\Delta\lambda > 0.02\,\mathrm{\AA}$ \\
   &          &	1133&498 & \multicolumn{2}{c}{}	&		    		       \\
   &          &	1135&324 & \multicolumn{2}{c}{}	&		    		       \\
   &          &	1135&588 & \multicolumn{2}{c}{}	&		    		       \\
   &          &	1136&603 & \multicolumn{2}{c}{}	&		    		       \\
   &          &	1136&986 & 1137&000	        &		    		       \\
   &          &	1137&625 & \multicolumn{2}{c}{}	&		    		       \\
   &          &	1138&248 & \multicolumn{2}{c}{}	&		    		       \\
   &          &	1138&497 & \multicolumn{2}{c}{}	&		    		       \\
   &          &	1139&278 & 1139&220	        & $\Delta\lambda > 0.02\,\mathrm{\AA}$ \newline
                                                  blend with \ion{Ge}{v}               \\
   &          &	1140&703 & \multicolumn{2}{c}{}	&		    		       \\
   &          &	1141&003 & 1141&015	        &                                      \\
   &          &	1141&095 & \multicolumn{2}{c}{}	&		    		       \\
   &          &	1141&344 & \multicolumn{2}{c}{}	&		    		       \\
   &          &	1142&792 & \multicolumn{2}{c}{}	&		    		       \\
   &          &	1142&925 & \multicolumn{2}{c}{}	&		    		       \\
   &          &	1142&938 & \multicolumn{2}{c}{}	& prev. unid. 	    		       \\
   &          &	1143&196 & \multicolumn{2}{c}{}	&		    		       \\
   &          &	1143&403 & \multicolumn{2}{c}{}	&		    		       \\
   &          &	1144&136 & 1144&160	        & $\Delta\lambda > 0.02\,\mathrm{\AA}$ \\
   &          &	1145&151 & \multicolumn{2}{c}{}	&		    		       \\
   &          &	1146&057 & \multicolumn{2}{c}{}	&		    		       \\
   &          &	1146&149 & \multicolumn{2}{c}{}	& too strong in model	    	       \\
   &          &	1147&020 & 1147&040             &                                      \\
   &          &	1147&371 & 1147&425             & $\Delta\lambda > 0.02\,\mathrm{\AA}$ \\
   &          &	1148&922 & 1148&915	        &		    		       \\
   &          &	1149&398 & 1149&370	        & $\Delta\lambda > 0.02\,\mathrm{\AA}$ \\
   &          &	1149&486 & \multicolumn{2}{c}{}	&		    		       \\
   &          &	1149&608 & \multicolumn{2}{c}{}	&		    		       \\
   &          &	1149&873 & 1149&855	        &		    		       \\
   &          &	1150&743 & \multicolumn{2}{c}{}	& 		    		       \\
   &          &	1151&368 & \multicolumn{2}{c}{}	&		    		       \\
   &          &	1152&985 & 1152&980             &  				       \\
   &          &	1153&160 & \multicolumn{2}{c}{}	&		    		       \\
   &          &	1155&027 & 1155&045             &  				       \\
   &          &	1155&725 & 1158&750	        &		    		       \\
   &          &	1156&520 & \multicolumn{2}{c}{}	&		    		       \\
   &          &	1156&885 & \multicolumn{2}{c}{}	&        	    		       \\
   &          &	1158&122 & \multicolumn{2}{c}{}	&		    		       \\
   &          &	1158&475 & \multicolumn{2}{c}{}	&		    		       \\
   &          &	1158&759 & 1158&750             &  				       \\
   &          &	1160&091 & \multicolumn{2}{c}{}	& weak	    		               \\
   &          &	1160&221 & \multicolumn{2}{c}{}	&		    		       \\
   &          &	1161&971 & \multicolumn{2}{c}{}	&		    		       \\
   &          &	1162&281 & \multicolumn{2}{c}{}	&		    		       \\
   &          &	1162&401 & \multicolumn{2}{c}{}	&		    		       \\
   &          &	1164&090 & \multicolumn{2}{c}{}	& blend of \newline \Ionww{Zn}{5}{1165.082, 1164.101} \\
   &          &	1164&632 & \multicolumn{2}{c}{}	& blend with \ion{O}{iv}               \\
   &          &	1165&189 & \multicolumn{2}{c}{}	&		    		       \\
   &          &	1165&706 & \multicolumn{2}{c}{}	& blend with \ion{C}{iii}	       \\
   &          &	1165&716 & \multicolumn{2}{c}{}	& blend with \ion{C}{iii}	       \\
   &          &	1165&880 & \multicolumn{2}{c}{}	& blend with \ion{Xe}{vi}	       \\
   &          &	1170&105 & \multicolumn{2}{c}{}	&		    		       \\
   &          &	1171&106 & 1171&130	    	&	                               \\
   &          &	1173&366 & \multicolumn{2}{c}{}	&		    		       \\
   &          &	1174&346 & 1174&325             &                                      \\	 
   &          &	1174&945 & \multicolumn{2}{c}{}	& blend with \ion{C}{iii}              \\
   &          & 1176&122 & \multicolumn{2}{c}{}	& blend with \ion{C}{iii}              \\
   &          &	1178&759 & \multicolumn{2}{c}{}	&		    		       \\
   &          &	1179&145 & \multicolumn{2}{c}{}	&		    		       \\
   &          &	1179&179 & \multicolumn{2}{c}{}	&		    		       \\
   &          &	1179&969 & 1180&005             & $\Delta\lambda > 0.02\,\mathrm{\AA}$ \\
   &          &	1180&018 & \multicolumn{2}{c}{}	& blend of \newline \Ionww{Zn}{5}{1180.018, 1180.025} \\
   &          &	1182&019 & \multicolumn{2}{c}{}	&		    		       \\
   &          &	1182&567 & \multicolumn{2}{c}{}	&		    		       \\
   &          &	1183&041 & \multicolumn{2}{c}{}	&		    		       \\
   &          &	1183&100 & \multicolumn{2}{c}{}	&		    		       \\
   &          &	1183&158 & \multicolumn{2}{c}{}	&		    		       \\
   &          &	1183&314 & \multicolumn{2}{c}{}	&		    		       \\
   &          &	1185&619 & \multicolumn{2}{c}{}	&		    		       \\
   &          &	1185&645 & \multicolumn{2}{c}{}	&		    		       \\
   &          &	1185&676 & \multicolumn{2}{c}{}	&		    		       \\
   &          &	1185&898 & 1185&905             &  				       \\
   &          &	1185&948 & 1185&955             & blend of \newline \Ionww{Zn}{5}{1185.948, 1185.961} \\
   &          &	1186&057 & \multicolumn{2}{c}{}	&	      			       \\
   &          &	1186&447 & 1186&420	        & $\Delta\lambda > 0.02\,\mathrm{\AA}$ \\
   &          &	1187&664 & \multicolumn{2}{c}{}	& too strong in model	               \\
   &          &	1187&706 & \multicolumn{2}{c}{} & too strong in model	               \\
\hline
\end{longtable}
\end{longtab}

\cite{dreizlerwerner1996} determined \Teffw{70\,000 \pm 4000} and \loggw{7.50 \pm 0.25} for 
\re.
This was recently verified by well-matched ionization equilibria
of Kr and Xe \citep[\ion{Kr}{vi}\,/\,\ion{Kr}{vii}, \ion{Xe}{vi}\,/\,\ion{Xe}{vii}]{werneretal2012}
and Ge \citep[\ion{Ge}{v}\,/\,\ion{Ge}{vi}]{rauchetal2012}.
We adopt these values and start our calculation based on the final model of 
\citet{werneretal2012} that considers opacities of He, C, N, O, Ge, Kr, Xe,
and of the iron-group elements (Ca - Ni). We follow the same strategy described in
Sect.\,\ref{sect:models}, but this time, we reduced the size of the Ge model atom
in the line-formation calculations.

\begin{figure}
   \resizebox{\hsize}{!}{\includegraphics{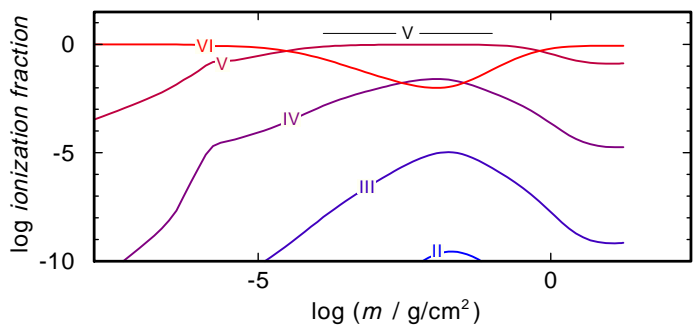}}
    \caption{Like Fig.\,\ref{fig:ionization}, for our \re model atmosphere.
            }
   \label{fig:ionization_re}
\end{figure}

The higher \Teff compared to that of \gb shifts the Zn ionization equilibrium strongly towards
higher ionization (Fig.\,\ref{fig:ionization_re}). \ion{Zn}{v} remains dominant while
\ion{Zn}{iv} is less occupied by a factor of about 100 at all depths.
The ionization fraction of \ion{Zn}{vi} is also much below that of \ion{Zn}{v} 
and we only expect weak lines. The strongest \ion{Zn}{vi} lines\footnote{\url{http://www.pa.uky.edu/~peter/atomic}} 
are located in the soft X-ray to EUV\footnote{extreme ultraviolet} wavelength range where we do not 
have high-quality observations to evaluate.

We determine a Zn abundance of
$2.7 \times 10^{-4} \pm 0.2\,\mathrm{dex}$ (about 155\,$\times$ solar)
to reproduce the observed \ion{Zn}{v} line profiles best (Fig.\,\ref{fig:znre0503}).

\section{Results and conclusions}
\label{sect:results}

The identified \ion{Zn}{iv} and \ion{Zn}{v} lines in the high-resolution UV spectra of \gb 
and \re are well reproduced with our newly calculated oscillator strengths by our
NLTE model-atmosphere calculations. 

We determined photospheric abundances of 
$\log\,\mathrm{Zn} = -5.52 \pm 0.2$ (mass fraction, $1.9 - 4.8\,\times\,10^{-6}$,  1.1 --   2.8 times the solar abundance) and
$\log\,\mathrm{Zn} = -3.57 \pm 0.2$                ($1.7 - 4.3\,\times\,10^{-4}$, 98   -- 248   times solar)
for 
the DA-type white dwarf \gb and 
the DO-type white dwarf \re, respectively.
The highly supersolar Zn abundance is in line with the high abundances of trans-iron elements
Ge \citep[650\,$\times$ solar,][]{rauchetal2012},
Kr (450\,$\times$ solar),
Xe \citep[3800\,$\times$ solar,][]{werneretal2012} in \re.

The identification of new lines due to trans-iron elements, e.g.,
Ga, Ge, As, Se, Kr, Mo, Sn, Te, I, and Xe \citep{werneretal2012}
and Zn \citep[and in this paper]{rauchetal2013} in \gb and \re promises to help 
enhance the understanding of extremely metal-rich white dwarf photospheres and their 
relation to AGB and post-AGB stellar evolution.
Their reproduction, 
i.e., the precise abundance determination, 
e.g., of 
Kr and Xe \citep{werneretal2012},
Ge and Sn \citep{rauchetal2012}, and
Zn (this paper) 
is strongly dependent on the available atomic data.
This remains a challenge for atomic and theoretical physicists.

\begin{acknowledgements}
TR is supported by the German Aerospace Center (DLR, grant 05\,OR\,0806).
Financial support from the Belgian FRS-FNRS is also acknowledged. 
PQ is research director of this organization.
This research has made use of the SIMBAD database, operated at CDS, Strasbourg, France.
Some of the data presented in this paper were obtained from the
Mikulski Archive for Space Telescopes (MAST). STScI is operated by the
Association of Universities for Research in Astronomy, Inc., under NASA
contract NAS5-26555. Support for MAST for non-HST data is provided by
the NASA Office of Space Science via grant NNX09AF08G and by other
grants and contracts. 

\end{acknowledgements}

\bibliographystyle{aa}
\bibliography{23491}

\begin{thebibliography}{10}
\expandafter\ifx\csname natexlab\endcsname\relax\def\natexlab#1{#1}\fi

\bibitem[{{Asplund} {et~al.}(2009){Asplund}, {Grevesse}, {Sauval}, \&
  {Scott}}]{asplundetal2009}
{Asplund}, M., {Grevesse}, N., {Sauval}, A.~J., \& {Scott}, P. 2009, \araa, 47,
  481

\bibitem[{{Cowan}(1981)}]{cowan1981}
{Cowan}, R.~D. 1981, {The theory of atomic structure and spectra} (Berkeley,
  CA, University of California Press)

\bibitem[{{Dreizler} \& {Werner}(1996)}]{dreizlerwerner1996}
{Dreizler}, S. \& {Werner}, K. 1996, \aap, 314, 217

\bibitem[{{Holberg} {et~al.}(1994){Holberg}, {Hubeny}, {Barstow}, {Lanz},
  {Sion}, \& {Tweedy}}]{holbergetal1994}
{Holberg}, J.~B., {Hubeny}, I., {Barstow}, M.~A., {et~al.} 1994, \apjl, 425,
  L105

\bibitem[{{Rauch} \& {Deetjen}(2003)}]{rauchdeetjen2003}
{Rauch}, T. \& {Deetjen}, J.~L. 2003, in Astronomical Society of the Pacific
  Conference Series, Vol. 288, Stellar Atmosphere Modeling, ed. I.~{Hubeny},
  D.~{Mihalas}, \& K.~{Werner}, 103

\bibitem[{{Rauch} {et~al.}(2012){Rauch}, {Werner}, {Bi{\'e}mont}, {Quinet}, \&
  {Kruk}}]{rauchetal2012}
{Rauch}, T., {Werner}, K., {Bi{\'e}mont}, {\'E}., {Quinet}, P., \& {Kruk},
  J.~W. 2012, \aap, 546, A55

\bibitem[{{Rauch} {et~al.}(2013){Rauch}, {Werner}, {Bohlin}, \&
  {Kruk}}]{rauchetal2013}
{Rauch}, T., {Werner}, K., {Bohlin}, R., \& {Kruk}, J.~W. 2013, \aap, 560, A106

\bibitem[{{Sugar} \& {Musgrove}(1995)}]{sugarmusgrove1995}
{Sugar}, J. \& {Musgrove}, A. 1995, Journal of Physical and Chemical Reference
  Data, 24, 1803

\bibitem[{{Werner} {et~al.}(2003){Werner}, {Deetjen}, {Dreizler}, {Nagel},
  {Rauch}, \& {Schuh}}]{werneretal2003}
{Werner}, K., {Deetjen}, J.~L., {Dreizler}, S., {et~al.} 2003, in Astronomical
  Society of the Pacific Conference Series, Vol. 288, Stellar Atmosphere
  Modeling, ed. I.~{Hubeny}, D.~{Mihalas}, \& K.~{Werner}, 31

\bibitem[{{Werner} {et~al.}(2012){Werner}, {Rauch}, {Ringat}, \&
  {Kruk}}]{werneretal2012}
{Werner}, K., {Rauch}, T., {Ringat}, E., \& {Kruk}, J.~W. 2012, \apjl, 753, L7

\end{thebibliography}

\end{document}